\documentclass[journal,transmag]{IEEEtran}

\usepackage[cmex10]{amsmath}
\usepackage[utf8]{inputenc}

\usepackage{graphicx}
\usepackage{amsmath}
\usepackage{enumitem}
\usepackage{bbm}

\usepackage{cuted}
\usepackage{lipsum}

\usepackage{multirow}

\usepackage{mathrsfs}
\usepackage{color,soul}
\usepackage{algorithm}
\usepackage{algpseudocode}
\usepackage{hyperref} 
\usepackage{amsmath,amssymb}
\usepackage{amsmath,amsfonts,amssymb}

\usepackage{amsthm}

\usepackage{eurosym}
\usepackage{subfig}

\usepackage{graphicx}
\makeatletter
\let\old@ps@headings\ps@headings
\let\old@ps@IEEEtitlepagestyle\ps@IEEEtitlepagestyle
\def\psccfooter#1{%
	\def\ps@headings{%
		\old@ps@headings%
		\def\@oddfoot{\strut\hfill#1\hfill\strut}%
		\def\@evenfoot{\strut\hfill#1\hfill\strut}%
	}%
	\def\ps@IEEEtitlepagestyle{%
		\old@ps@IEEEtitlepagestyle%
		\def\@oddfoot{\strut\hfill#1\hfill\strut}%
		\def\@evenfoot{\strut\hfill#1\hfill\strut}%
	}%
	\ps@headings%
}
\makeatother

\graphicspath{{sections/chPhase/}}
\begin{document}

%
\title{\Huge{Towards Phase Balancing using Energy Storage}}

 \author{\IEEEauthorblockN{Md Umar Hashmi\IEEEauthorrefmark{1},
 José Horta\IEEEauthorrefmark{2},
 Lucas Pereira\IEEEauthorrefmark{3}, 
 Zachary Lee\IEEEauthorrefmark{4},
 Ana Bu\v{s}i\'c\IEEEauthorrefmark{1}, and
 Daniel Kofman\IEEEauthorrefmark{2}}
 \IEEEauthorblockA{\IEEEauthorrefmark{1} INRIA and the Computer Science Dept. of Ecole Normale Sup\'erieure, CNRS, PSL Research University,
 	Paris, France}
 \IEEEauthorblockA{\IEEEauthorrefmark{2} Laboratory of Information, Networking and Communication Sciences, Télécom ParisTech,
 Paris, France}
 \IEEEauthorblockA{\IEEEauthorrefmark{3} ITI, LARSyS, Técnico Lisboa and prsma.com,
 Funchal, Portugal}
 \IEEEauthorblockA{\IEEEauthorrefmark{4}Electrical Engineering Department, California Institute of Technology,
 Passdena, CA, USA}
 }

\maketitle

\begin{abstract}
Ad-hoc growth of single-phase-connected distributed energy resources, such as solar generation and electric vehicles, can lead to network unbalance with negative consequences on the quality and efficiency of electricity supply. 
Case-studies are presented for a substation in Madeira, Portugal and an EV charging facility in Pasadena, California. These case studies show that phase imbalance can happen due to a large amount of distributed generation (DG) and electric vehicle (EV) integration.
We conducted stylized load-flow analysis on a radial distribution network using an openDSS-based simulator to understand such negative effects of phase imbalance on neutral and phase conductor losses, and in voltage drop/rise. We evaluate the integration of storage in the distribution network as a possible solution for mitigating effects caused by imbalance. 
We present control architectures of storage operation for phase balancing. 
Numerically we show that relatively small-sized storage (compared to unbalance magnitude) can significantly reduce network imbalance. We identify the end node of the feeder as the best location to install storage.
\end{abstract}
\section{Introduction} 
Most of the transmission and distribution of electrical energy is done in three phases. However,  low voltage  distribution of electricity is often performed as if the three phase system was a set of three independent single phase lines.
In a perfect scenario, the load on three phases with the same line length would be completely balanced, however, in real-world, asymmetries in each of the three phases are bound to happen due to unbalanced loading and cable length.
Network unbalance describes a condition in a poly-phase system in which the phasors of voltage or current are not equal in magnitude and/or the phase angle between consecutive phasors are not all equal. 
Network unbalance can have negative impacts in quality and efficiency of electricity supply and in long term can lead to a number of problems including thermal aging, equipment life reduction and derating of the capacity of induction machines.

%
The localized injections of DG energy could cause over-voltage problems at a localized level leading to false tripping of circuit breakers and relays. 
Furthermore, intermittent RES can cause load balancing problem \cite{hashmi2018effect} at central level and also among phases \cite{r4liu2014probabilistic}.
Against this background, this paper presents an empirical exploration of the effects of unbalanced generation and load on three phase radial distribution network. 
To state more concretely, we perform load-flow analysis on a radial distribution network with unbalanced phases by connecting single phase RES and EV loads on a 3 phase 4 wire distribution system.
We explore the system unbalance in terms of (a) line and neutral losses, (b) voltage unbalance factor (VUF) and (c) voltage rise/drop. 
We observe network unbalance is affected by: (i) the location of network where single phase loads or generation is connected, (ii) neutral and phase losses are more pronounced compared to VUF for a compact radial distribution network like densely populated cities, 
(iii) sparse or congested networks with significant voltage drop can lead to substantial increase in VUF.

We provide results of two unbalance case studies, in a substation in Madeira, Portugal, and in an EV charging facility in Pasadena California. 
In the first case study we detail power network operators’ current empirical approach for planning customer phase allocations to reduce network unbalance. However, in spite of a careful analysis before acceptance of DG installations and a static phase allocation, substation operators observed imbalance in the distribution side and in some cases are constrained to decline of new DG installations, hindering the development of RES in Madeira. In the second case, we observe experimentally that a majority of EVs connected to any one phase can unbalance the three-phase network. 
Both cases motivate the need to 
introduce storage (or load flexibility) for phase balancing.
We present storage based control architectures for achieving phase balancing by compensating active and reactive power between phases.
Finally, we also show a small size storage, compared to imbalance magnitude, can contribute to phase balancing noticeably.

This paper is divided into 7 sections. Section~\ref{phaseSection1} provides an introduction of phase unbalance. 
Section~\ref{phaseSection4} and Section~\ref{caltechcase} we describe case-studies for Madeira and Pasadena respectively showing imbalances in distribution networks due to DGs and EVs.
In Section~\ref{phaseSection2} we perform OpenDSS based radial distribution network simulations for identifying the effect of connecting single phase DG/EV in 3-phase system. 
Section~\ref{phaseSection3} present storage architectures and stylized storage control for phase balancing.
Section~\ref{phaseSection5} concludes the paper.
\section{Phase Balancing} 
\label{phaseSection1}
A three phase system have unbalanced voltage if the rms value of phase voltages\footnote{RMS or root mean square voltage is $V_{rms} = \sqrt{(V_a^2 + V_b^2 + V_c^2)/3}$.} are not the same and/or the phase angle between voltage phases are not exactly 120 degrees \cite{r9beharrysingh2014phase}, \cite{r7bchydro}.
Fortescue in 1918 developed symmetrical components for representation of any set of unbalanced phasors \cite{fortescue1918method}:
(i) a direct or \textit{positive} sequence in order (abc),
(ii) an inverse or \textit{negative} sequence in order (acb),
(iii) a homopolar or \textit{zero} sequence system in same direction. 

Current unbalance is related to voltage unbalance through network impedances. If the network impedances are asymmetric then voltage unbalance can occur even though the currents are perfectly balanced. 
In this work, we balance power in each phase leading to reduction in unbalance in current and voltage. 
The remaining part of this section we discuss the reasons, effects and mitigation of unbalance. 

\subsection{Cause of unbalance in three-phase power network}
\label{causesimbalancephase}
Asymmetries in each of the three phases can be because of unbalanced loading and cable length. Here we highlight the unbalance which could be caused  of large scale EV deployment and DG installation.
The connection of single phase EVs and DGs are random and often clustered in a certain area. 
In addition both EVs and PVs tend to be active in a synchronized manner, i.e., majority of the EVs are getting charged in the evening when people reach their homes after work and PVs generate when it is sunny. Such a synchronized operation of these loads and generation aggravates the problems for distribution system  operators who are obliged to ensure power quality at all times.
Electric vehicles require high charging current and longer periods of charging. 
EVs with 30kWh battery stores as much as the average US residence consumes in a day, making it significant portion of total household load \cite{r11fitzgerald2016electric}.
Furthermore, often EV charging is single phase which could cause voltage unbalance \cite{r12putrus2009impact}. 
The sale of EV will be upto 50\% of all new cars by 2030 \cite{r11fitzgerald2016electric}. 
Authors in \cite{r12putrus2009impact} note that voltage imbalance caused by EVs is unlikely to exceed the prescribed limits set by the utility provided EVs are reasonably distributed among three phases.
This could have adverse effect on voltage unbalance of a three-phase distribution network \cite{r15shahnia2013predicting}.

DG is intermittent, interfaced via single phase converter and solar PV also have a high impedance and low short circuit current making it more prone to cause unbalance in the power network \cite{r4liu2014probabilistic}, \cite{driesen2002voltage}.
Within Europe the voltage characteristic standard, EN50160 \cite{cenelec200150160}, states that the 10-min rms voltage should be between 90\% and 110\% of the nominal voltage most of the time and between 85\% and 110\% all of the time. The distribution feeders are designed such that the voltage magnitude becomes lower when moving along the feeder. Integrating DGs at distribution level makes the design condition of feeders invalid as over-voltages could occur due to localized power injections \cite{r3bollen2008integration}. The introduction of DG also changes the fault currents and will increase the risk of an incorrect protection operation.
Further, DG intermittency can aggravate the 3-phase imbalance.
For example solar generation can reduce by 70\% in few seconds due to passing clouds \cite{crabtree2011integrating}.
Thus, larger integration of DGs and PVs will increase the chances of unbalance in 3-phase network. 

\subsection{Effect of unbalance in three-phase power network}
Voltage unbalance can create a current unbalance 6-10 times the magnitude of voltage unbalance \cite{x1pge}.
Negative sequence for synchronous generator causes overheating in the inner construction of damper windings \cite{r4liu2014probabilistic}.
For transformers, zero sequence (one third of neutral current for 4-wire 3-phase system) flow could inverse the temperature of windings and cause parasitic losses in the transformer structure \cite{driesen2002voltage}.
It is estimated that there are 2.4 GWh, equivalent to \$134,000 additional annual losses due to presence of unbalance based on 17,600 transformers in Brazil \cite{salustiano2013unbalanced}.
Due to phase unbalance, share of positive sequence decreases leading to
reduction in capacity for carrying positive sequence current \cite{chen1995evaluation, czarnecki1995power}.
Phase unbalance can, therefore, limit the power transferred on a feeder \cite{r24wang2013phase}. 
Unbalance could also lead to preventive breaker or relay tripping and shut-down of a feeder \cite{r24wang2013phase}.
The voltage unbalance of low-voltage feeder may be seen by other feeders fed from the same distribution transformer. Furthermore, current unbalance can be propagated through the distribution transformer to the high voltage network \cite{r9beharrysingh2014phase},\cite{chindrics2007propagation}. 
\begin{table}[!htbp]
	\caption {Voltage Unbalance Norms} 
	\label{vuflimits}
	\begin{center}
		\begin{tabular}{| c | c|}
			\hline
			Utility/Standard & VUF Limit\\
			\hline
			PG\&E \cite{x1pge} & 2.5\%\\
			\hline
			NEMA MG-1-1988 \cite{nema}  & 1\%\\
			\hline
			BC Hydro - Standard Unbalance \cite{r7bchydro} & 2\%\\
			BC Hydro - Rural Unbalance \cite{r7bchydro} & 3\%\\
			\hline
			Europe EN 50160 - LV and MV \cite{euunbalance} & 2\%\\
			Europe EN 50160 - High Voltage \cite{euunbalance} & 1\%\\		
			\hline
		\end{tabular}
		\hfill\
	\end{center}
\end{table}	

\subsection{Mitigation of Unbalance in three-phase power network} 
Unbalance in power distribution network can be eliminated by design if the transmission lines are fully transposed and load is divided symmetrically in all phases, both of which are difficult to realize in low voltage (LV) distribution networks.
Due to unbalanced loading and unequal feeder lengths, phase imbalance exists. Power utilities aim to contain the unbalance within pre-decided norms, rather than completely eliminating the unbalance. Table~\ref{vuflimits} lists the norms derived from standards and utilities for voltage unbalance. Note the norms often are more detailed and are specific to the network and feeder. For instance in a radial distribution network, feeders away from the source have higher potential to cause unbalance than those close to the source. 
Many traditional solutions are available in literature for solving unbalance problem in three phase networks. Inclusion of shunt or series connected compensators are widely used traditionally. Recent work present feeder reconfiguration for phase balancing \cite{r24wang2013phase}.
Authors in \cite{r17wang2017phase} propose coordination of operation of data-center and DERs to reduce phase unbalance.
Authors in \cite{r25weckx2015load} propose load balancing with EV chargers and PV inverters in unbalanced distribution grids by adaptively selecting the phase based on unbalance.
Authors in \cite{r14lico2015phase} present phase balancing using EVs interfaced to the power network via a single phase connection which adaptively selects the phase to connect in the LV Danish grid.
Authors in \cite{horta2017market} identified the negative impact of local energy markets on voltage unbalance and in \cite{horta2018augmenting} used solid state switches to achieve phase balancing by dynamically allocating households to phases based on their local market commitments. Authors in \cite{sun2016phase} used energy storage batteries owned privately for performing arbitrage and phase balance considering battery degradation.
By balancing the voltage it helps in saving energy and money by increasing motor efficiency, transmission line capacity is better utilized, false trigger of protective relays are avoided and also prevents downtime due to motor failures.

Energy storage interfaced via a converter is capable of supplying active and reactive power for phase balancing of three phase network \cite{hashmi2019arbitrage}. Authors in \cite{chua2012energy} use storage for mitigating unbalance which could be caused by building integrated PV. 
\cite{r28chua2011mitigation} mitigates voltage unbalance in LV distribution network with high penetration of PV system using energy storage. The controller minimizes the current flow in the neutral line and experimental study indicates improved VUF. 

\section{Case-Study : LV Substation in Madeira Island}
\label{phaseSection4}
{This section presents a real data-based case study of phase imbalance in context of Madeira Island in Portugal. First, an overview of the local grid is provided, in order to give some context to the reader. Second, the case of a specific distribution substation in Madeira is presented, in order to motivate the need to improved grid control through the introduction of energy storage at the distribution station level.} 
\subsection{Low Voltage Distribution Substation} 
{The selected LV distribution substation is located in \textit{Estreito da Calheta}, one of the most south-western villages in Madeira Island, Portugal.  Madeira is a total energy island, where a DSO/TSO is responsible for the activities related to production, transport, distribution and commercialization of electric energy, including private micro-generation \cite{hashmi2019energy}.}

\begin{figure}[bh]
	\centering
	\includegraphics[width=0.3\textwidth]{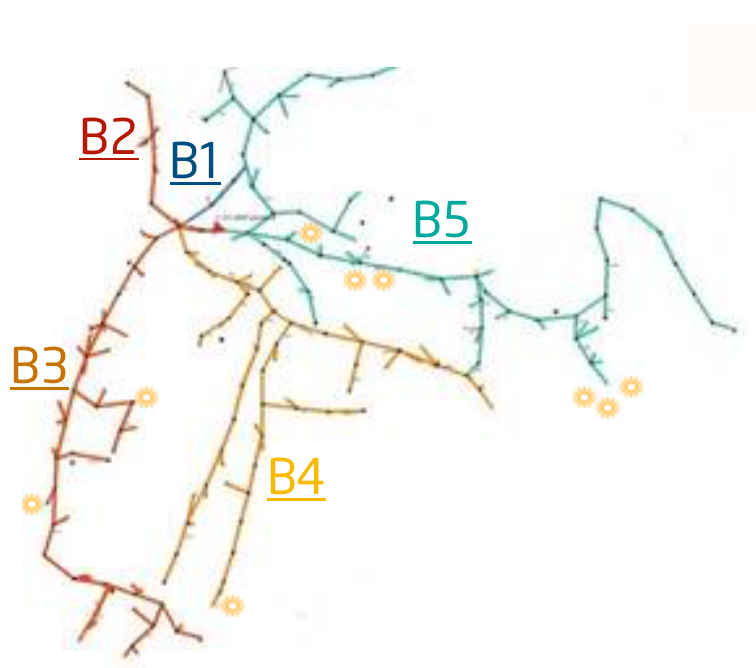} 
	\caption{Madeira substation with five output branches and the nine connected DGs; yellow denotes the location of DGs}
	\label{fig:ds_outputs}
\end{figure}

{This substation has a transformer with an apparent power of 250 kVA, connected in \textit{delta-wye}, which transforms voltage from the transmission grid (6600 V) to the distribution grid (400 V). The feeder capacity is of 62.5 kVA, i.e., 25\% of the transformer capacity. The daily average load is of 27 kW, with an average off-peak power of 1 kW, and peak power of 63 kW.}
{This substation is one of Madeira's LV substations with high amount of PV generation, with a total capacity of 36 kWp (14\% of the feeder capacity), distributed over nine DGs, three of which have three-phase installations (see Table~\ref{tab:upps}).} The distribution network is shown in Fig.~\ref{fig:ds_outputs}. 

The island power networks are more vulnerable to fluctuations compared to mainland power networks. This is because islands cannot have any help unlike mainland grid which can share resources. Due to this the power utility in Madeira is more conservative in installing new DGs in the distribution network. In order to contain the effects of DG, DSO/TSO encourages the DG owners to only self-consume the locally generated renewable generation. This is done to ensure voltage stability in the distribution network. DG injections can cause voltage surges in sparse distribution networks. Due to this norm, the DG growth in Madeira is hindered with average installation in Madeira is below 0.6 kWp. With more flexible norms, we expect the renewable generation growth in Madeira to explore with year-round sunshine and high wind speeds.
As such, there is still a lot of room to connect new DGs and loads, which will require additional efforts to keep the grid properly balanced.
\begin{table}[th]
	\scriptsize
	\centering
	\caption{DGs at the substation, and their installed capacities.}
	\label{tab:upps}
	\begin{tabular}{|c|c|c|c|}
		\hline
		\textbf{DG} & \textbf{ kWp} & \textbf{Branch} & \textbf{Phases}\\
		\hline
		1 & 5.17, 3.3, 1.95 & B5 & A B C \\
		4 & 3.45 & B3 & A\\
		5 & 3.45 & B3 & B \\
		6 & 3.45 & B4 & C\\
		7 & 5.17 & B5 & A\\
		8 & 5.17 & B5 & B\\
		9 & 5.17 & B5 & C\\
		\hline
	\end{tabular}
\end{table}

\begin{figure}[ht]
	\centering
	\includegraphics[width=3.5in]{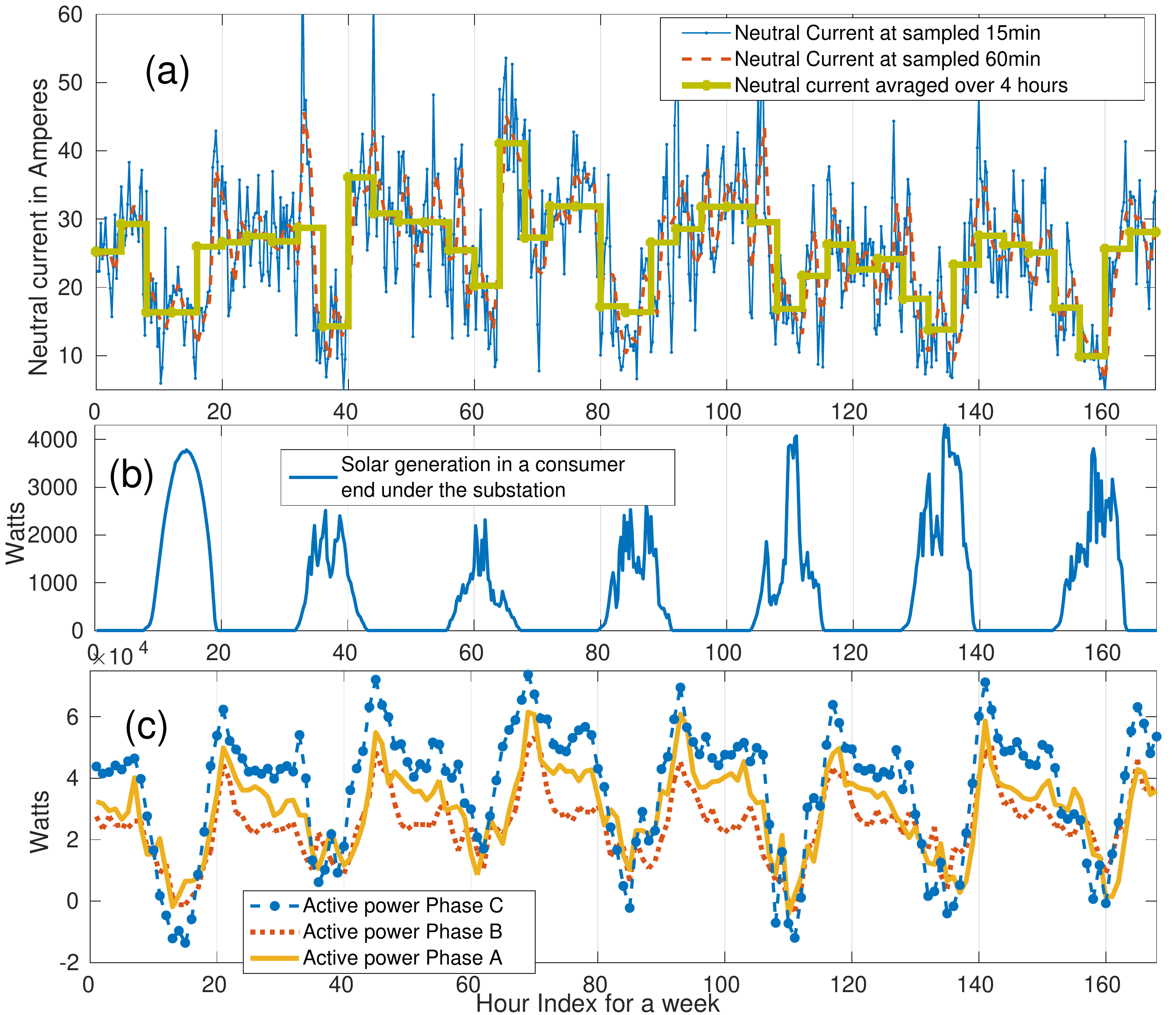} 
	\caption{\small{(top) Neutral current measured at substation, (center) solar generation in a consumer end fed from the sub-station, and (bottom) active power supplied from the sub-station.}}
	\label{fig:imbalanceMadeira}
\end{figure}
\begin{figure}[ht]
	\centering
	\includegraphics[width=3.5in]{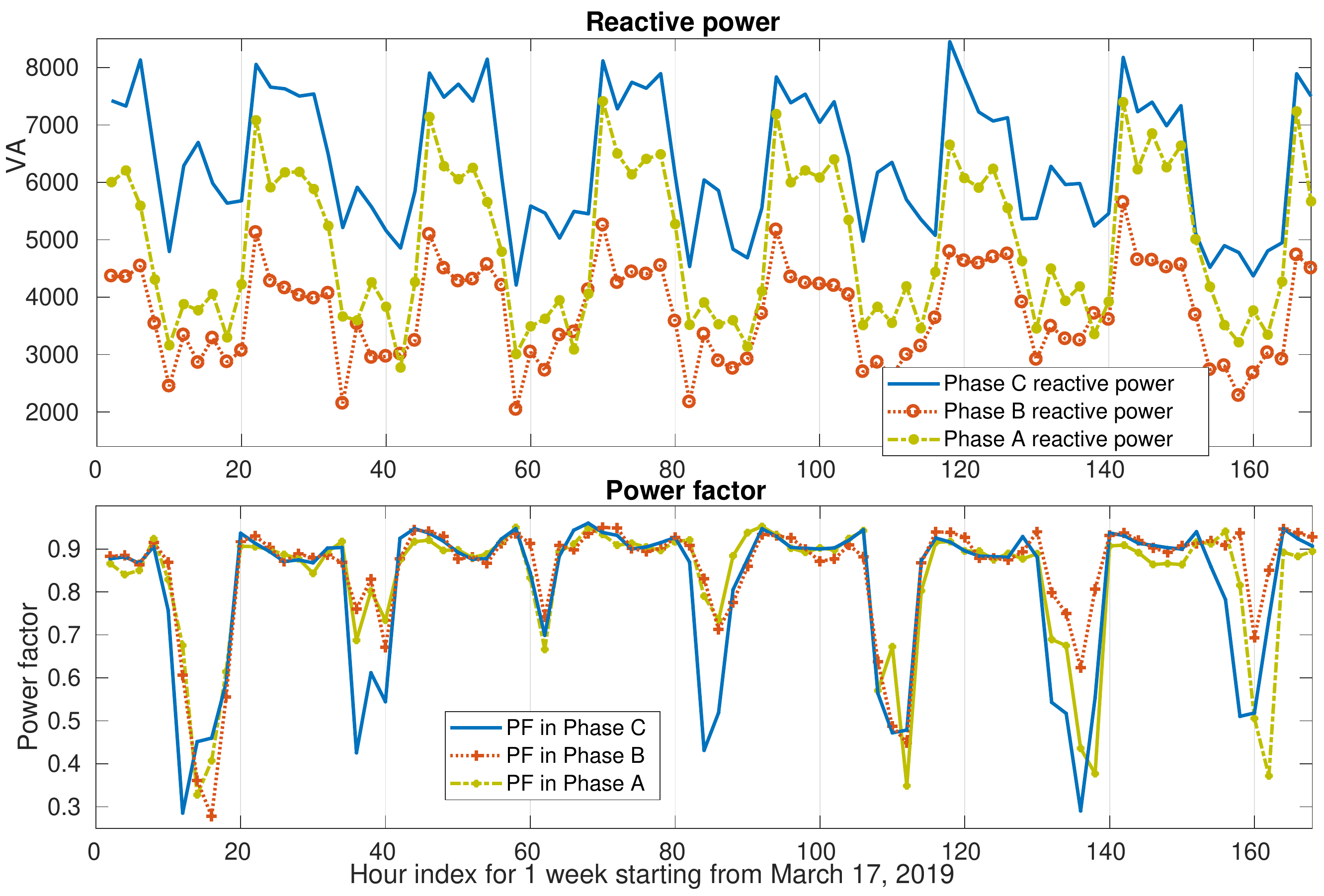} 
	\caption{\small{(top) reactive power supplied by each phase, and (bottom) power factor seen by the sub-station.}}
	\label{fig:pfMadeira}
\end{figure}
\vspace{-15pt}

{\subsection{Examples of Unbalance} 
	Fig.~\ref{fig:ds_outputs} and Table~\ref{tab:upps} shows the Madeira electric grid is empirically balanced, i.e., the DSO/TSO relies on the experience of the distribution team to carefully plan and manage the grid, by always trying to do the best distribution of the installations in each phase (thus reducing the phase unbalance occurrence), and having the adequate extensions of the conductors (thus avoiding voltage fluctuations and the edge of the grid). 
}

{Still, and despite the best efforts, due to the stochastic nature of energy consumption and renewable generation, there is still some considerable phase unbalance. 
	For example, in Fig.~\ref{fig:imbalanceMadeira} 
	we observe that while there is solar PV production, the unbalance is stable across the three phases, which greatly contrast to the periods without solar PV production where the unbalance is significantly higher.
	{Observe that the network is nearly balanced for lightly loaded condition during the day when solar production is maximum and the neutral current is minimum. The network is designed to be balanced when DG production is maximum. However, for evening peaks the neutral is almost 100\% greater compared to during solar peak generation during the day.} The plots are based on real-data collected for a week from March 17, 2019.
	%
	%
}

Further, in Fig.~\ref{fig:pfMadeira} the reactive power imbalance is shown. Note that the power factor and reactive power deviates significantly during the day leading to noticeable difference among the phases. {To summarize, during the day when solar is generating, the active power among phases are more balanced but reactive power and power factor is more unbalanced.}

{Against this background, it is very relevant to study the possibility of providing improved grid control through the introduction of energy storage at the distribution station level. Ultimately, the ability to coordinate active and reactive power balances at the distribution level would represent a major step towards safely increasing the injection of renewable energy sources in the Madeira electric grid. Furthermore, since many of the rural LV substations in Madeira Island share the same characteristics (i.e., low number of consumers, and mostly single-phase installations), there is a high replication potential for such a solution.}

\section{Case-Study : EV Charging in Pasadena} 
\label{caltechcase}
Phase unbalance can be an issue behind the meter, particularly in large-scale EV charging facilities such as those in workplaces or public parking facilities. To demonstrate this, we consider the Caltech Adaptive Charging Network (ACN) located in a parking garage in Pasadena, CA \cite{lee2018large}. In this section, we first provide an overview of the charging facility, then demonstrate how significant current unbalance can occur due to normal charging behavior. This highlights the need for phase balancing techniques which can account not only for instantaneous unbalances, but also long term unbalances caused by differences in the energy demand on each phase.

\subsection{Electric vehicle charging}
{EVs are expected to dominate future transportation. Charging EV batteries can place a massive load on the local electrical infrastructure.
In Table~\ref{tab:evchar} we list some EVs and their battery characteristics. The batteries can either be charged using three-phase or single-phase AC connection. In the case of AC level-2 charging, a single-phase connection is used by the AC/DC converter inside the EV. In the case of DC charging, a single-phase or three-phase connection can be used to feed an AC/DC converter outside the EV, which then feeds DC current directly to the EV's battery. Table~\ref{tab:evcharginglevels} lists the standard single and three-phase EVSEs (charging ports) and their rated power transfer capability. For instance, Nissan Leaf can be completely charged within 4 hours using 1-phase 32A charging EVSE. 
All charging points in Caltech ACN Testbed are of single phase and 32A rated type. 
}
\begin{table}[th]
	\scriptsize
	\caption{EV battery characteristics \cite{evch2}}
	\label{tab:evchar}
	\begin{center}
	\begin{tabular}{|p{1.5cm}|p{2.2cm}|p{1.3cm}|p{2.3cm}|}
		\hline
		EV make & Warranty & Battery & Charge times\\
		\hline
		Nissan Leaf & 8yrs./100,000 miles & 30 kWh & 8h at 230V AC, 15A,\\
		Chevrolet Bolt & 8yrs./100,000 miles & 60 kWh & 10h at 230V AC, 30A \\
		Tesla model S & 8yrs./unlimited miles &70, 90 kWh & 9h with 10kW charger \\
		\hline
	\end{tabular}
	\end{center}
\end{table}

\begin{table}[th]
	\scriptsize
	\centering
	\caption{EV charging socket characteristics \cite{evch1}}
	\label{tab:evcharginglevels}
	\begin{center}
		\begin{tabular}{|p{2.6cm}|p{1.8cm}|p{2.5cm}|}
			\hline
			Charger type (230V AC) & Rated power & Time to charge 30kWh\\
			\hline
			1-$\phi$ 16 A & 3.7 kW & 8 hours\\
			1-$\phi$ 32 A & 7.4 kW & 4 hours \\
			1-$\phi$ 16A/$\phi$ & 11kW & 2h 45 min \\
			3-$\phi$ 32A/$\phi$ & 22 kW & 1h 22 min \\
			\hline
		\end{tabular}
	\end{center}
\end{table}

\subsection{The Caltech ACN Testbed} 
The ACN testbed at Caltech has delivered over 846 MWh of electricity to charge electric vehicles since early 2016. In this study, we will consider a subset of the ACN, which consists of 54 single-phase, level-2 EVSEs  connected line-to-line, each having a maximum charging power of 6.6 kW. Power is supplied to the EVSEs via a three-phase network at 208 V\textsubscript{LL}, which is provided by a 150 kVA delta-wye transformer. Originally, the network was designed with 12 EVSEs on AB, 14 on BC, and 14 on CA. However, early in the project, it was decided to replace two existing EVSEs, both of which happened to be on AB, with pods of 8 EVSEs each, resulting in 26 EVSEs on AB. This unequal allocation of EVSEs only exacerbates the unbalances that naturally occur due to randomness in user charging behaviors.

\subsection{Examples of Unbalance} 
\begin{figure}[h]
    \centering
    \includegraphics[width=3.4in]{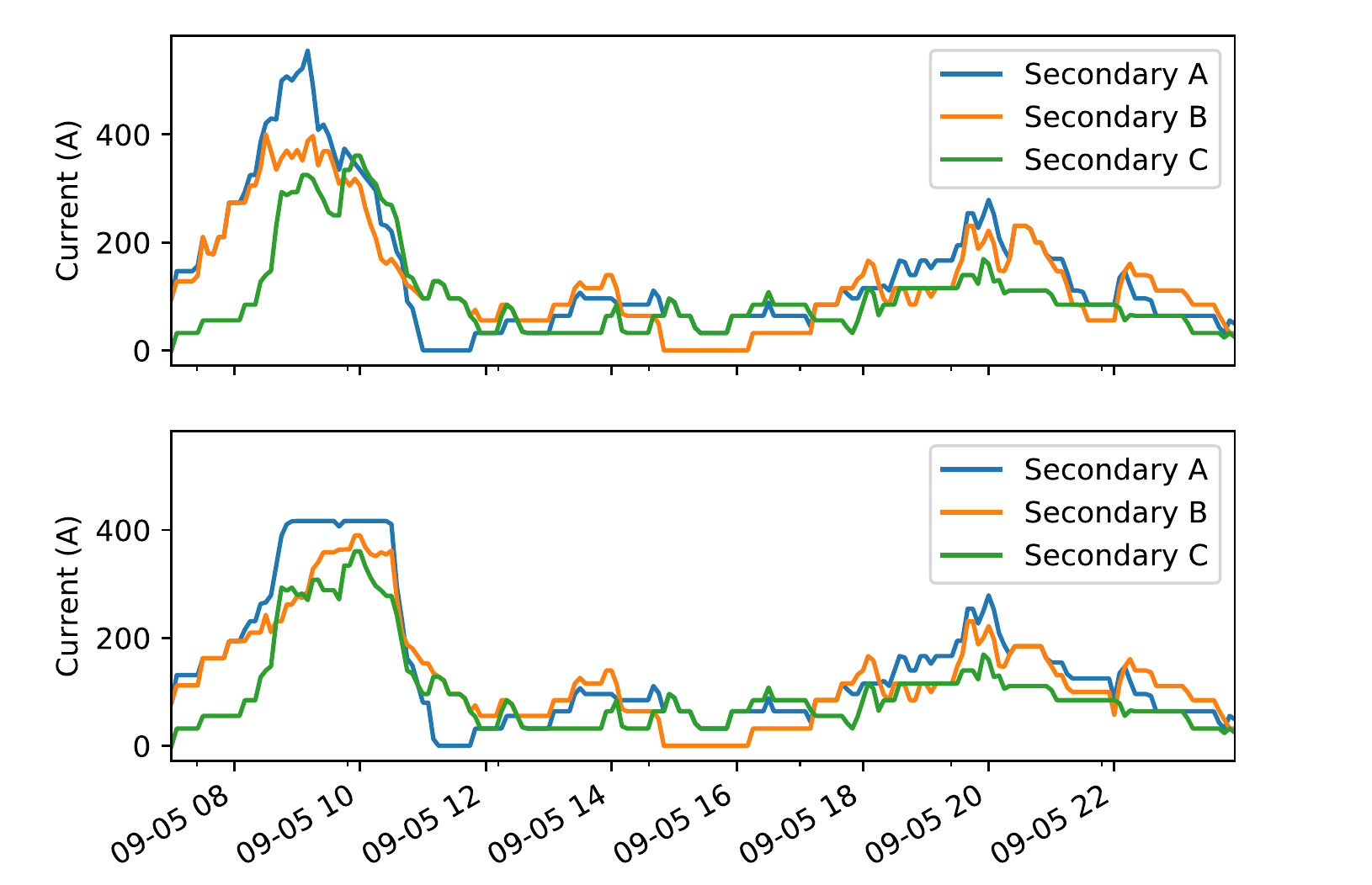} 
    \caption[Phase current imbalance for uncontrolled EV charging at the Caltech ACN.]{\small{Current unbalance for uncontrolled EV charging (top) and smart EV charging (bottom) at the Caltech ACN. Both plots are simulated based on real data collected Sept. 5, 2018.}}
    \label{fig:ev_unbalance}
\end{figure}

To demonstrate the unbalance present in the Caltech ACN, we consider data collected from the ACN on Sept. 5, 2018 \cite{lee2019ACN-Data}, \cite{lee2019datasite}. While the ACN currently uses smart charging algorithms to prevent overloads of system components, most charging facilities provide uncontrolled charging, so we present the current unbalance for both cases. In order to simulate charging activities and line currents within the ACN, we utilize ACN-Sim \cite{lee2019ACN-Sim}. The top panel of Fig.~\ref{fig:ev_unbalance} shows the current unbalance that results from uncontrolled charging, while the bottom panel shows the unbalance from the smart charging algorithm used in the actual ACN \cite{lee2018large}. In both cases, current unbalance can be significant, differing by as much as 280 A between lines A and C in the uncontrolled case. 

One reason for this significant unbalance is that the total energy needed on each phase can be quite different. For example, on Sept. 5, 2019, the total energy demand on EVSEs on phase AB was 408 kWh, while BC was 178 kWh, and CA was 232 kWh. Because of these unbalanced energy demands, balancing currents requires us to distribute load not only in time but also between phases, something smart charging alone cannot accomplish. This shortcoming of smart charging approaches motivates us to look for new ways to accomplish phase balancing in large-scale charging facilities. Doing so will allow us to increase charging capacity by better utilizing existing infrastructure as well as reduce transformer wear caused by current unbalance. 
Energy storage can be one of the ways in which phase balancing can be performed. In the subsequent section, we analyze the effects of phase imbalance on power quality and losses in the network using simulations on a radial distribution network. 

\section{Phase imbalance simulations for radial distribution network} 
\label{phaseSection2}
In this section, we perform phase unbalance simulations on a radial distribution network by connecting single phase renewable generation and electric vehicle loads on a three phase four wire distribution system.
We identify that system network imbalance indicators as (a) losses in the neutral, (b) line losses and (c) VUF. 
We observe that effect on system unbalance is affected by: (i) the location of network where single phase loads or generation is connected in the network, (ii) the effect of connecting single phase load and generation is nearly symmetrical for VUF, however, the effect of neutral and line losses are not symmetrical for loads and generation, (iii) networks which are sparse or congested with significant voltage drop could lead to significant VUF values. 

The system considered is shown in Fig.~\ref{simbase}.
The nominal case has a balanced load in each phases as we aim to understand the variation caused due to integration of DGs/EVs in one of the phase.
We perform a sensitivity analysis by placing different levels of RES (0\%, 10\%, 20\%, ..., 90\%, 100\%, 120\% of max load in each phase) at points N1 (close to feeder) and N5 (furtherest to feeder). 
 In this experiment we assume the worst-case condition where all these single phase DGs/EVs are connected to one of the phase.
For evaluation we are interested in observing the variations of following parameters at nodes:
(a) VUF: for each node; we use International Electrotechnical Commission definition of VUF (= the ratio of the magnitudes of negative sequence over positive sequence)  \cite{pillay2001definitions}, 
(b) per-unit voltage: for each phase at each node, 
(c) active power: for each phase at each node, 
(d) reactive power: for each phase at each node, 
(e) losses incurred in each of the phases and 
(f) losses incurred in the neutral conductor.

\begin{figure}[!htbp]
	\centering
	\includegraphics[scale=0.31]{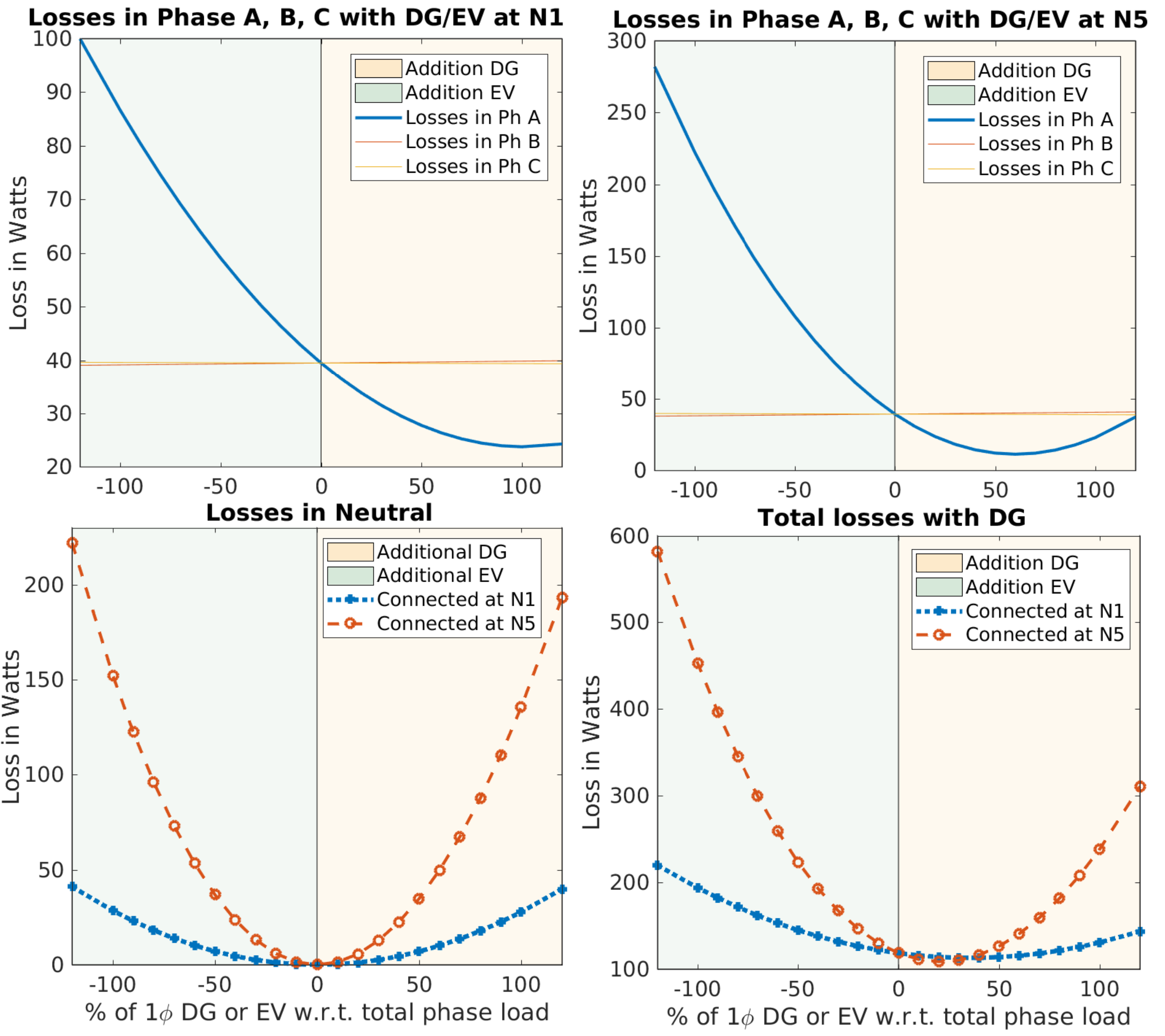} 
	\caption{\small
		{Losses with: (i) DG/EV at N1 (ii) DG/EV at N5}
	}
	\label{figloss}
\end{figure}

\begin{figure}[!htbp]
	\centering
	\includegraphics[scale=0.28]{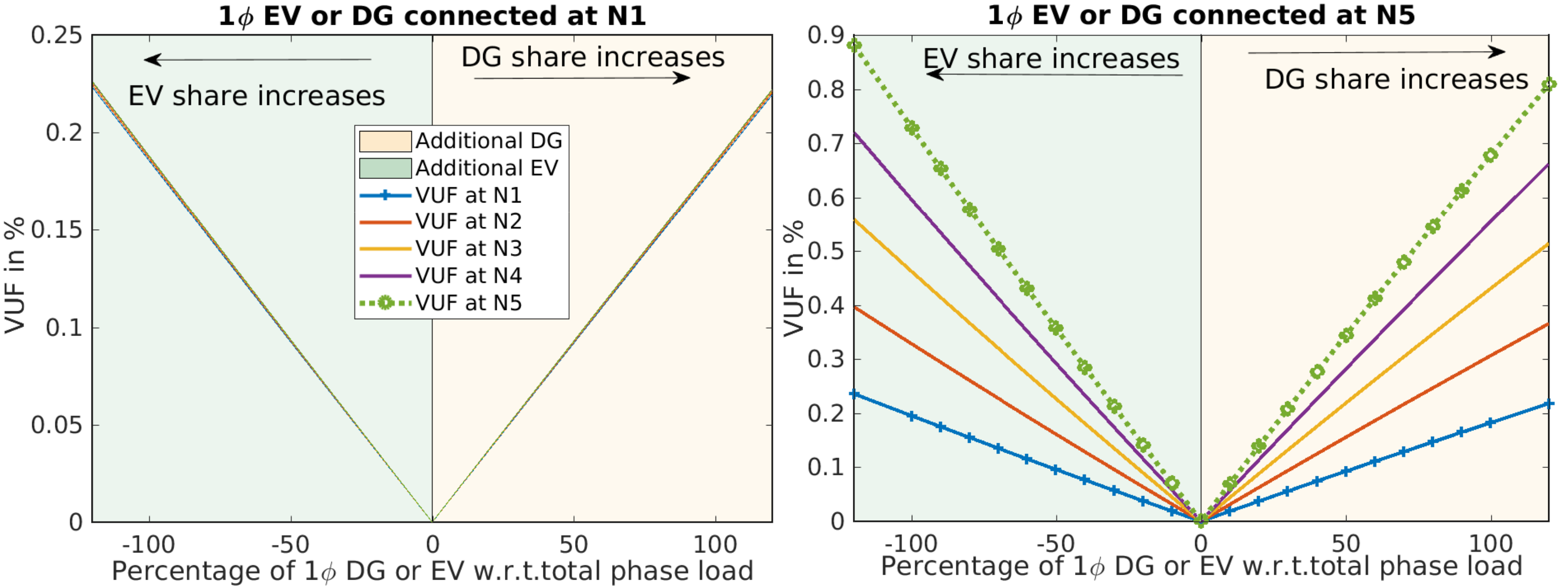} 
	\caption{\small{VUF with (i) DG/EV at N1 and (ii) DG/EV at N5}
	}
	\label{figvuf}
\end{figure}

\begin{figure}[!htbp]
	\centering
	\includegraphics[scale=0.32]{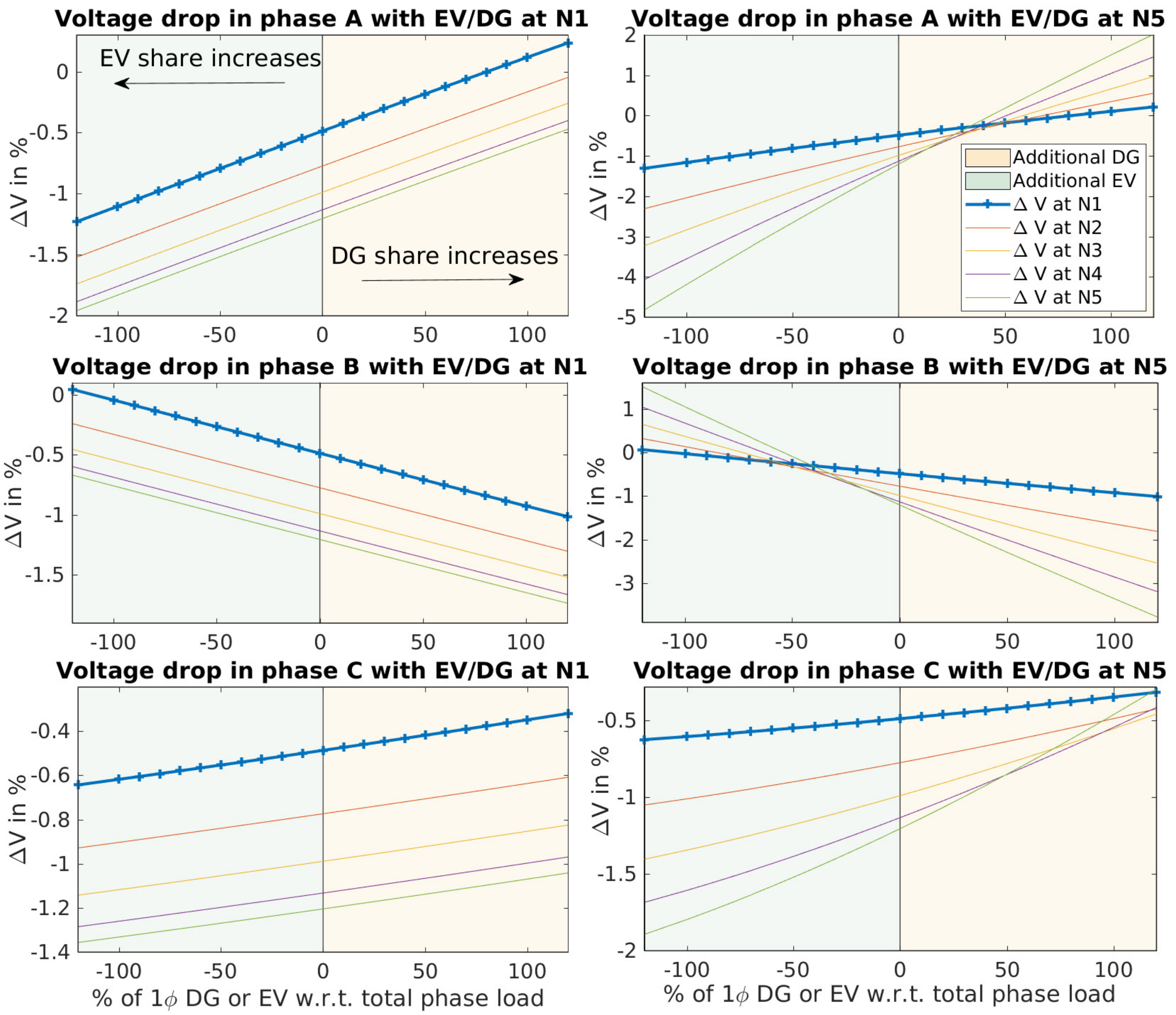} 
	\caption{\small
		{Voltage drop with (i) DG/EV at N1 (ii) DG/EV at N5}
	}
	\label{figdrop}
\end{figure}

The key observations using simulations for a radial network shown Fig.~\ref{simbase} are as follows:
%
\begin{itemize}
	\item VUF is not affected until the lines have significant voltage drop due to high resistance or overloaded and/or network is sparse with significant line losses. For a compact network with low drop in voltage with respect to voltage at the generation feeder in a radial distribution network, VUF is not significant even for a large share of DGs/EVs connected to only one phase, refer to Fig.~\ref{figvuf}. 
	From Table~\ref{vuflimits} we observe the  VUF limit lies
	within 1-3\%. For compact network the VUF rises to less than 0.22\% for 120\% (compared to phase load) of DGs/EVs connected to N1 (refer to Table~\ref{basecase}) and 0.9\% for DGs/EVs connected to N5 (refer to Table~\ref{basecasen5}). However, VUF is a crucial index for networks which are either congested and/or sparse, refer to Table~\ref{overloadn5} to Table~\ref{sparsen5} (marked in red). 
	\item For single phase DG connected close to the feeder has a near to uniform effect on VUF compared to DG connected farther away which affects the distant nodes much more than nodes closer to the feeder. Fig.~\ref{figvuf} shows that increase in share of DG connected at N5 almost linearly increases the VUF. 
	\item Contrary to prevalent notion that adding renewables helps in reducing voltage drop if connected at distant nodes, however, DG not balanced along the phases could reduce the losses in a phases (note Phase A losses in Fig.~\ref{figloss}) but increases the losses in the neutral conductor drastically. Thus the total losses in effect are still large for large share of renewables. However, with increase integration of EVs the losses increase in phase and neutral without any ambiguity. 
	%
	\item The last plot of Fig.~\ref{figloss} shows that DGs can be designed to reduce the total loss in the distribution network. These results further improve if we assume DG placement to be balanced in each phase. 
	\begin{figure}[!htbp]
		\center
		\includegraphics[width=2.7in]{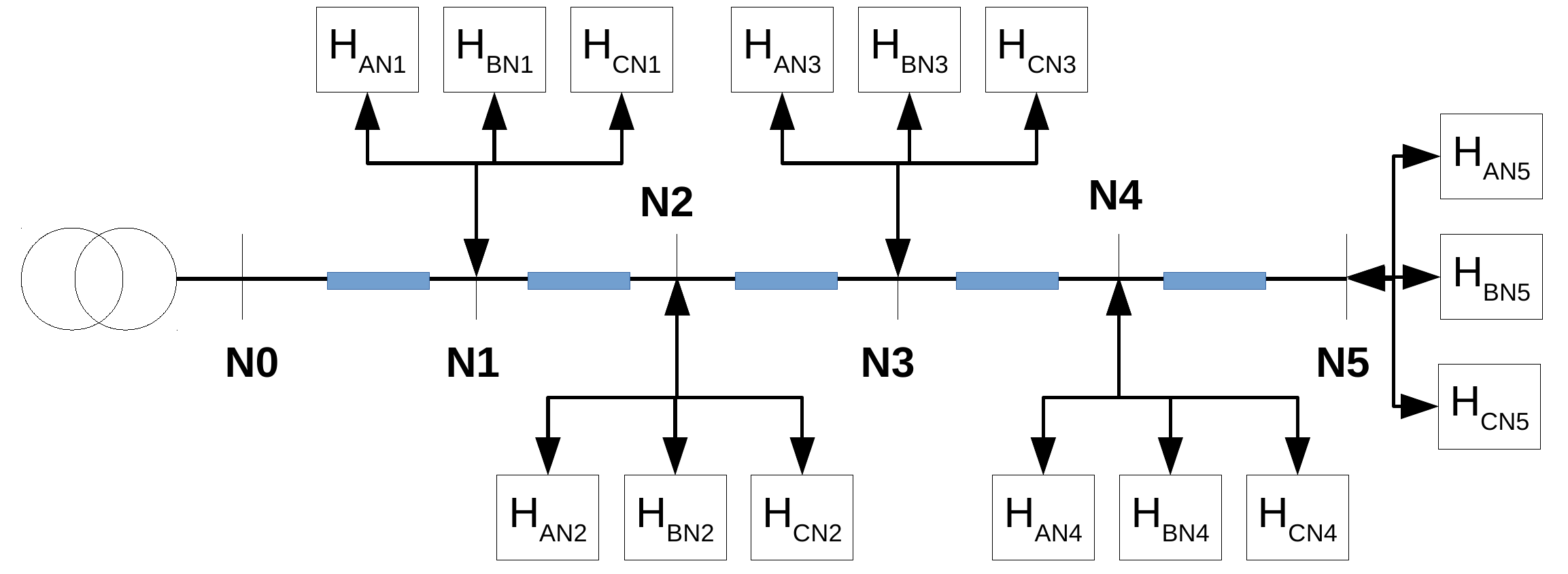} 
		\caption{Simulation Baseline Model}\label{simbase}
	\end{figure}
	\item Considerable increase (almost 3 to 4 times) voltage drop in the injecting phase reduces with DG connected at N5 compared to DG connected at N1. However, this configuration increases the voltage drop in a phase. 
	\item Fig.~\ref{figloss} shows the total phase losses, neutral losses and total losses for DGs/EVs connected at N1 and N5. The total losses for DG/EV connected at the distant feeder are significantly higher compared to the case where the DG/EV integration close to the feeder. 
	\item Fig.~\ref{figdrop} shows the voltage drop in percentage for integration of DG/EV at N1 and N5. The drop in voltage decreases in the phase where DG is connected and increases when EV is connected.
\end{itemize}
\begin{table}[!htbp]
	\scriptsize
	\caption {5kW total phase load with DG or EV at N1} \label{basecase}
	\begin{center}
		\begin{tabular}{p{2.3cm}p{0.5cm}p{0.5cm}p{0.6cm}p{0.6cm}p{0.9cm}p{0.9cm}} 
			\hline 
			EV/DG at N1& Mean VUF& Max VUF &Neutral Losses (kWh)& Total Phase Losses & Sum of Voltage drop  N1&Sum of Voltage drop  N5 \\
			\hline 
			\hline 
			Nominal Case & 7e-8	&1e-7&	0&	1.97&	-1.46&	-3.61\\
			40\% Balanced DG& 7e-8&	1e-7 &	0 &	1.48&	-1.10&	-3.25\\
			40\% Balanced EV& 9e-8&	2e-7&	0	&2.72&	-1.83&	-3.98\\
			\hline
			120\% Unbalanced DG& 0.221&	0.222&	\textbf{0.664}&	1.73&	-1.098& -3.25\\
			120\% Unbalanced EV& 0.225&	0.226&	\textbf{0.684}&	2.98&	-1.83&	-3.98\\
			\hline
		\end{tabular}
	\end{center}
\end{table}
\begin{table}[!htbp]
	\scriptsize
	\caption {5kW total phase load with DG or EV at N5} \label{basecasen5}
	\begin{center}
		\begin{tabular}{p{2.3cm}p{0.5cm}p{0.5cm}p{0.6cm}p{0.6cm}p{0.9cm}p{0.9cm}} 
			\hline 
			EV/DG at N1& Mean VUF& Max VUF &Neutral Losses (kWh)& Total Phase Losses & Sum of Voltage drop  N1&Sum of Voltage drop  N5 \\
			\hline 
			Nominal Balance Case & 8e-8	&1e-7&	0&	1.97&	-1.46&	-3.61\\
			Balanced + 40\% DG& 6e-8&	1e-7	&0&	0.73&	-1.09&	-2.06\\
			Balanced + 40\% EV& 8e-8&	1e-7	&0&	4.50&	-1.84&	-5.20\\
			\hline
			120\% Unbalanced DG& 0.514&	0.809&	\textcolor{red}{3.22}&	1.96&	-1.11&	-2.06\\
			120\% Unbalanced EV& 0.559&	0.881&	\textcolor{red}{3.70}&	6.00&	-1.87&	-5.21\\
			\hline
		\end{tabular}
	\end{center}
\end{table}
\vspace{-10pt}
\begin{table}[!htbp]
	\scriptsize
	\caption {50kW phase load (\textit{Overload}) with DG/EV at N5} \label{overloadn5}
	\begin{center}
		\begin{tabular}{p{2.3cm}p{0.5cm}p{0.5cm}p{0.6cm}p{0.6cm}p{0.9cm}p{0.9cm}} 
			\hline 
			EV/DG at N1& Mean VUF& Max VUF &Neutral Losses (kWh)& Total Phase Losses & Sum of Voltage drop  N1&Sum of Voltage drop  N5 \\
			\hline 
			\hline 
			Balanced + 40\% DG& 1e-7&	1e-7&	0&	83.1&	-11.64&	-22.06\\
			Balanced + 40\% EV& 1e-7&	2e-7&	0&	679.2&	-23.86&	-66.84\\
			\hline
			120\% Unbalanced DG& \textcolor{red}{5.24}&	\textcolor{red}{8.11}&	368.5&	267.7&	-14.53& -24.27\\
			120\% Unbalanced EV& \textcolor{red}{4.79}&	\textcolor{red}{7.21}&	116.8&	380.9&	-18.38&	-44.37\\
			\hline
		\end{tabular}
	\end{center}
\end{table}
\begin{table}[!htbp]
	\scriptsize
	\caption {5kW phase load (\textit{Sparse}) with DG/EV at N5}
	\label{sparsen5}
	\begin{center}
		\begin{tabular}{p{2.3cm}p{0.5cm}p{0.5cm}p{0.6cm}p{0.6cm}p{0.9cm}p{0.9cm}} 
			\hline 
			EV/DG at N1& Mean VUF& Max VUF &Neutral Losses (kWh)& Total Phase Losses & Sum of Voltage drop  N1&Sum of Voltage drop  N5 \\
			\hline 
			\hline 
			Balanced + 40\% DG& 8e-8&	1e-7&	0&	8.1	&-8.46&	-18.71\\
			Balanced + 40\% EV& 1e-7	&2e-7&	0	&63.8	&-17.49	&-58.99\\
			\hline
			120\% Unbalanced DG& \textcolor{red}{4.46}&	\textcolor{red}{7.21}&	3.7& 26.2&	-10.44&	-20.24\\
			120\% Unbalanced EV& \textcolor{red}{4.35}&	\textcolor{red}{6.81}&	1.3&	39.5&	-13.90&	-40.03\\
			\hline
		\end{tabular}
	\end{center}
\end{table}

\section{Energy storage for phase balancing}

\label{phaseSection3}


In previous sections we highlighted the negative impact of phase unbalance in the efficiency and quality of electricity supply, particularly in the case of massive deployment of distributed energy resources, both through stylized simulations and real case studies. In this section we analyze the usage of storage by the Distribution System Operator (DSO) as a solution to such issues by balancing active and reactive power among phases. We present possible architectures for the connection of storage resources to phases, relying in some cases on phase selectors (such as solid state switches) for the dynamic allocation of storage resources, and we discuss the impact of their size and location along the feeder based on stylized simulation results.

Using storage as a solution to phase unbalance has several advantages with respect to the traditional approach of switching the phase connection of individual households manually, based on historical data, as well as to more recent approaches, such as dynamically switching phase connections using solid state switches and individual households forecasts (or commitments in local energy markets \cite{horta2018augmenting}). For instance, fine-grained dynamic control over the distribution of active and reactive power among phases can be achieved; \cite{kisacikoglu2011reactive} note that the reactive power output of a battery
does not affect its State-of-Charge (SoC) and is governed by active power output and converter rating \cite{hashmi2019arbitrage}. Another beneficial aspect is that control algorithms may depend on the aggregated load/production forecast rather than on less precise individual household forecasts. Furthermore, storage can provide additional services for the DSO, such as load shifting to reduce congestion.


\subsection{Architectures of Storage Solutions}

\begin{figure}[!htbp]

\center

\includegraphics[width=3.6in]{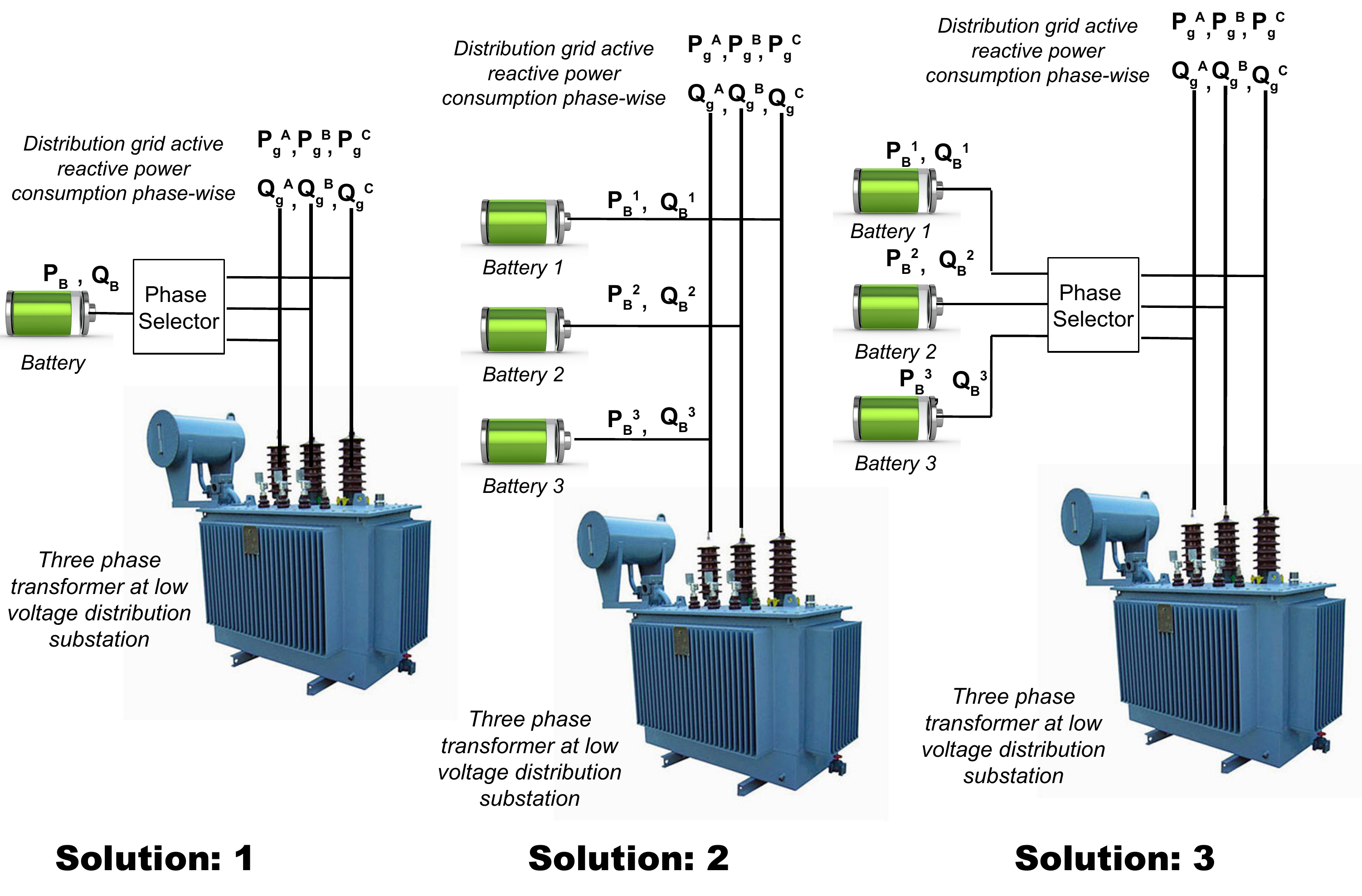}

\caption[Storage control architectures used at distribution level for balancing active and reactive power among the three phases.]{\small{Storage control architectures used at distribution level for balancing active and reactive power among the three phases \cite{hashmi2019optimization}.} }\label{phasebalancing}

\end{figure}

We briefly introduce three storage control architectures for phase balancing depicted in Figure~\ref{phasebalancing} and subsequently rely on stylized simulations to evaluate several aspects relevant for their applicability in real world, such as size and location.




\textit{Architecture 1 - one storage device and phase selector}:

A single battery is used to compensate imbalance among three-phases. A phase selection algorithm would define the phase to which the battery is connected at each time slot and a control strategy would define storage charging/discharging schedule to minimize the imbalance.













\textit{Architecture 2 - three storage devices each one dedicated to a phase}:
Three dedicated storage devices are connected each one to a specific phase, but a controller would define a coordinated storage charging/discharging schedule to minimize the imbalance. For instance, at some point in time, the battery at the most loaded phase would discharge while simultaneously the battery at the least loaded phase would charge. If the aggregated load at each point in time should not be modified (in case load shifting by DSO is not allowed),  a constraint can be added for the accumulated charged and discharged energy at every time slot to sum 0.

%













\textit{Architecture 3 - three storage devices and a phase selector for each storage}: In this architecture, batteries can be dynamically assigned to phases depending among other things on their current SoC and on the adjustment required on the distribution of load/generation among phases so as losses are reduced and power quality is maintained. The flexibility of this architecture can better accommodate for probable future variations, both on balancing requirements as well as on additional services, such as reducing curtailment of renewable energies. For example, consider one of the phases has an increase in installed DGs, then greater charging capacity may be needed during sun hours, otherwise part of such generation would be curtailed for ensuring voltage, current and imbalance are within limits.





Architecture 1 implies that, in addition to balancing load among phases, storage will be shifting demand and/or production in time. Although this is positive for the DSO, it may not be allowed by regulation or it may not be always economically feasible. The second architecture gives the possibility to the DSO to better balance load among phases without shifting load in time, having no impact on the aggregated load seen by the transformer at each timeslot. The limitation of this architecture is that batteries are statically connected to a certain phase, which reduces the space of solutions for the design of a control algorithm. This limitation can be avoided by using the third architecture with the possibility to dynamically allocate batteries to phases according to their SoC, the number of cycles, etc. Furthermore, the third architecture is a superset of the other two. In the following section we will use the first and the second architecture to analyze the potential impact of storage and provide hints for design of control algorithms.

\subsection{{Phase Balancing with Storage: Stylized Example}}



In simulations in Section \ref{phaseSection2}, we observe that for a radial distribution network placing DG or EV  at the farthest node can lead to higher imbalance compared to connecting the same amount of DG or EV at a node closer to the feeder. This indicates that storage placed at the farthest end can provide a maximum amount of correction, as storage can either act as a generator or load. We present a stylized example for showing significantly small-sized storage can substantially correct phase imbalance. 

\begin{figure}[!htbp]

\centering

\includegraphics[width=3.2in]{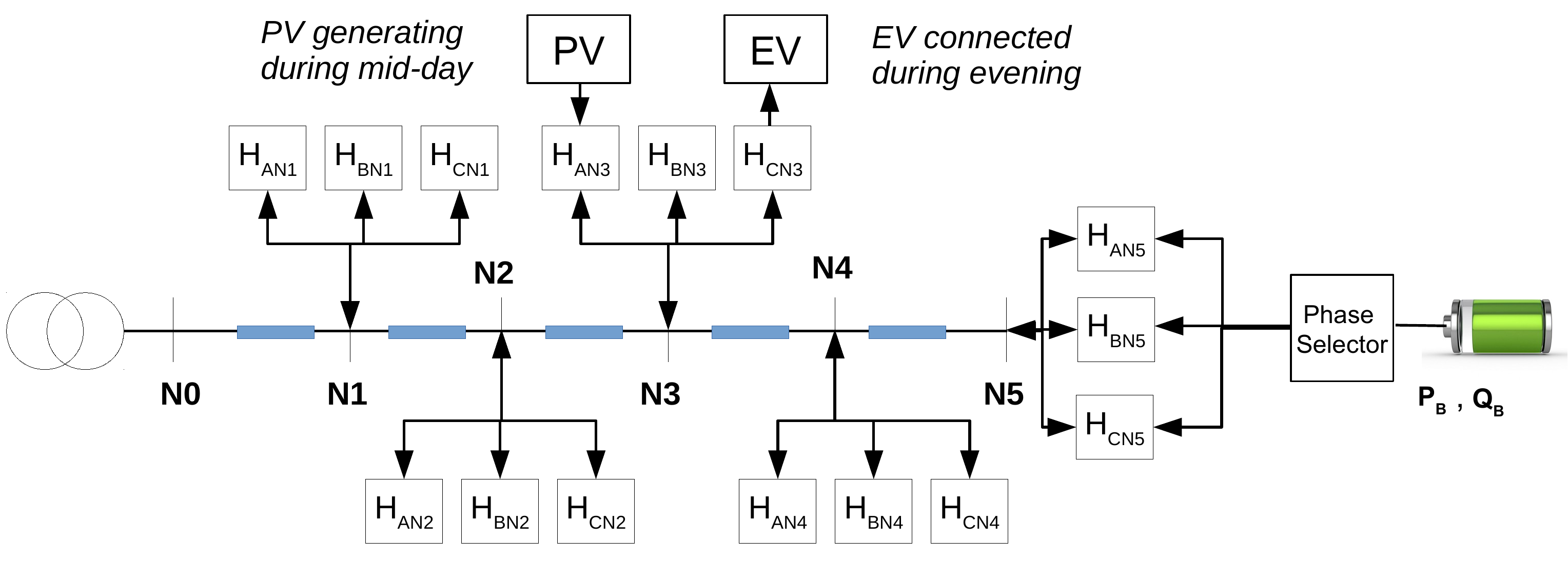} 

\caption{Simulation scenario for the stylized example.}\label{sysImbalance}

\end{figure} 


Fig.~\ref{sysImbalance} shows an example of the system considered for the simulations for the case with architecture 1 and the storage connected at node 5. In general, each phase under balanced condition have 10 kW of load with each consumer having 2 kW load, while for the unbalanced condition 10 kW\footnote{We refer to this value as the imbalance magnitude. For a more precise analysis this concept would be defined differently, for instance as the maximum of the mean squared distance among the active power of the phases. } of solar generation is connected on phase A between 10:00 and 15:00 and 10 kW of additional load (EVs) are connected on the same phase A\footnote{EVs and DGs are put on the same phase only for the sake of clarity.} from 18:00 to 23:00.

For the test of architecture 1 we consider a battery of size\footnote{we refer to power capacity, while we assume an energy capacity enough to charge or discharge for 5 hours at the corresponding power.} 3 kW (a third of the imbalance magnitude) connected to phase A charging from 10:00 to 15:00 and discharging from 18:00 to 23:00. If the EVs where connected to another phase the battery phase selector would be operated.

For the test of architecture 2 we consider 3 batteries of size 1kW each, and two control scenarios, one in which the effect of phase balancing is combined with load shifting, and the other in which only phase balancing is performed. In both cases the battery connected to phase A charges from 10:00 to 15:00 and then discharges from 18:00 to 23:00, while the other two batteries in phases B and C discharge from 10:00 to 15:00 and charge from 18:00 to 23:00. The difference in the second scenario is a constraint that limits the aggregate energy charged and discharged at each time to "sum to 0" so that the batteries do not shift load in time. 

\begin{figure}[!htbp]

\centering

\includegraphics[width=3.0in]{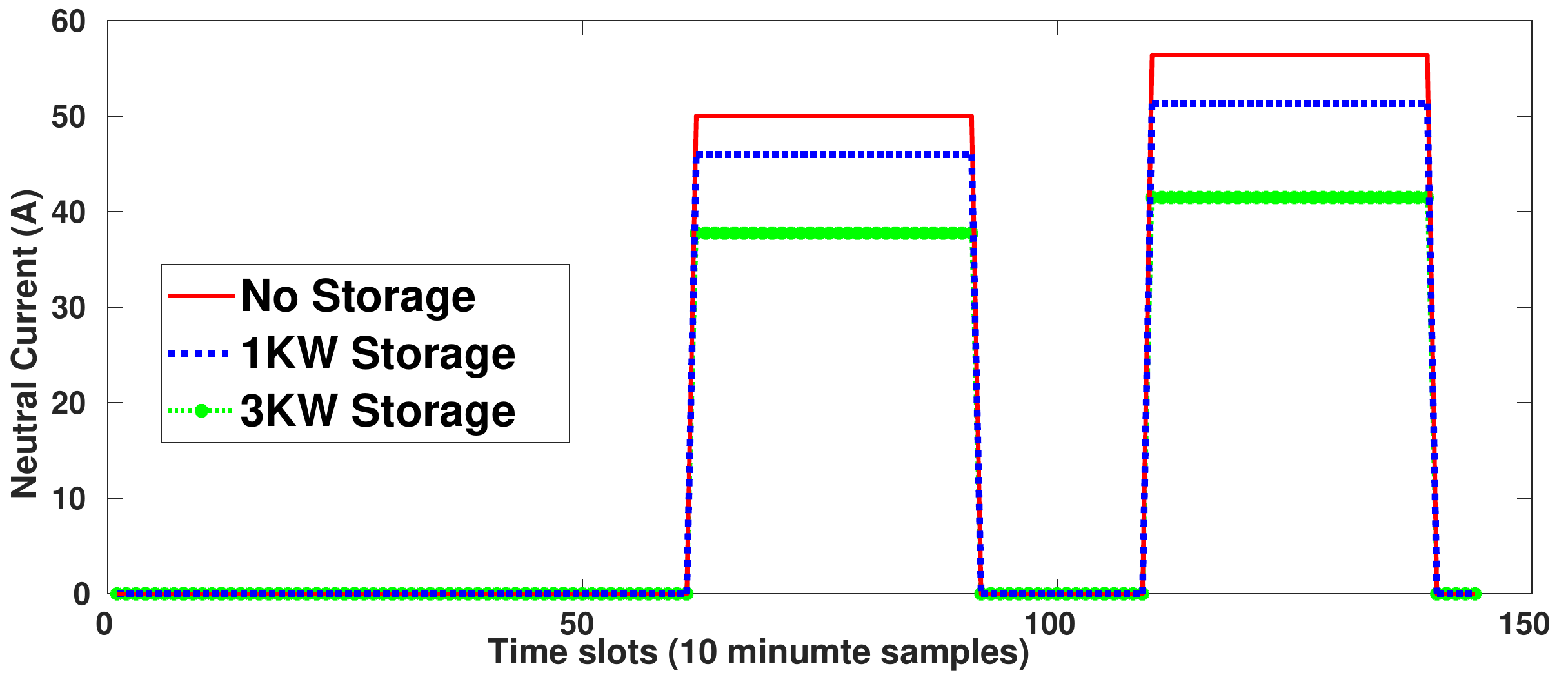} 

\caption{Impact on neutral current.}\label{Ilinetoy}

\end{figure}

\textit{Results and discussion}:
The results obtained show that the action of storage can significantly reduce the imbalance metrics identified previously. We will discuss first the results of the test for architecture 1 in terms of neutral currents, voltage deviations and voltage unbalance.
With respect to the impact on neutral current, Fig.~\ref{Ilinetoy} shows the variations in time for the case without storage and with two battery sizes, 1kW and 3kW, located at N5. We can see that these variations can be clearly associated to periods with phase unbalance, showing as well how a comparable small size battery can considerably reduce the neutral current, which has a corresponding quadratic reduction in losses.

\begin{figure}[!htbp]

\centering

\includegraphics[width=3.4in]{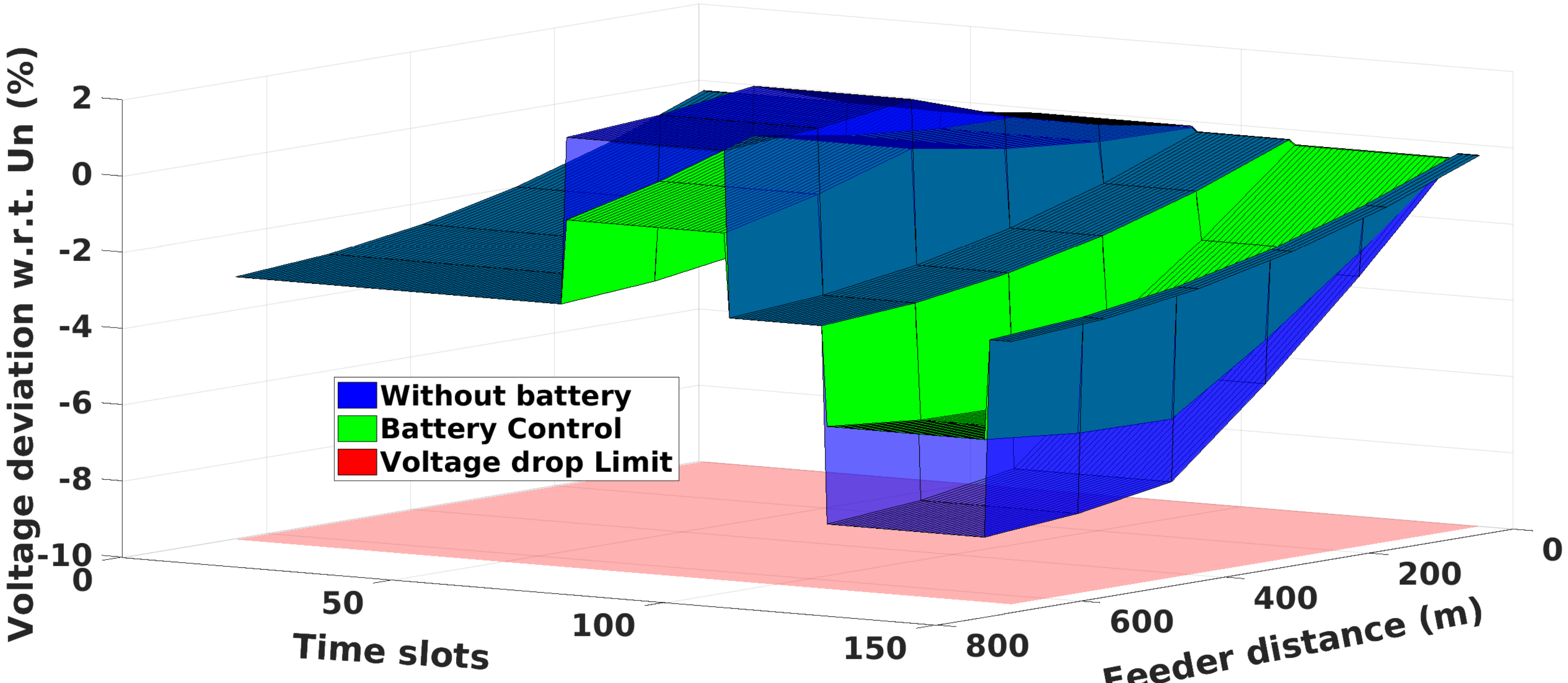} 

\caption{Impact on voltage deviation.}\label{Vdroptoy}

\end{figure}

\begin{figure}[!htbp]

\centering

\includegraphics[width=3.4in]{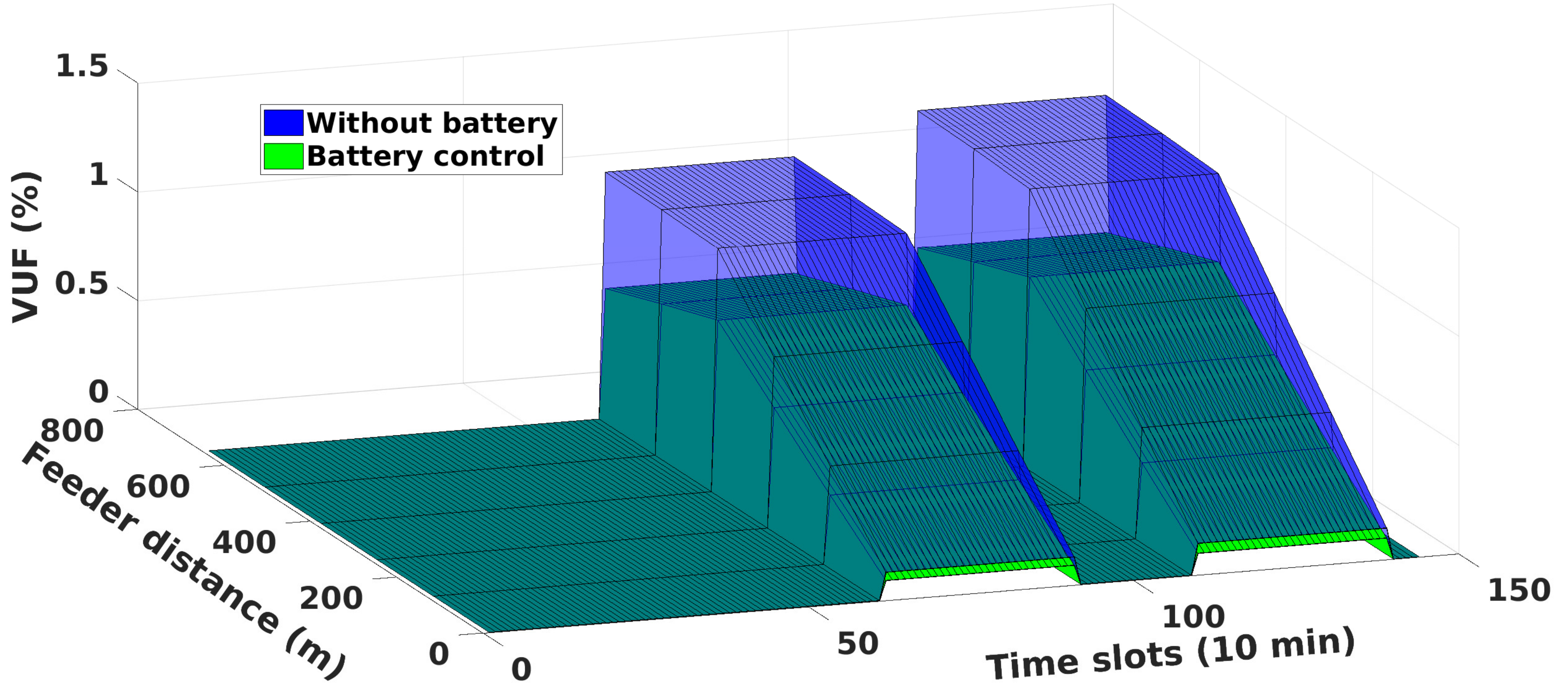} 

\caption{Impact on voltage unbalance factor.}\label{VUFtoy}

\end{figure}

Effect of storage inclusion on voltage deviations and VUF is shown in Fig.~\ref{Vdroptoy} and Fig.~\ref{VUFtoy} respectively.
We observe that the voltage deviations and VUF increases as we go farther away from the feeder.
effect along the time axis for the two unbalanced periods and as expected the effect of phase unbalance gets worst as voltage drops with the distance from the transformer and as it drops/rises around the unbalance source at node 3. In both cases the compensation effect of a 3kW storage is considerable, in particular keeping voltage deviations within the permissible voltage drop limit and reducing VUF to safer values of around 1\%.

For the test of architecture 2 we present a table showing the relevance of the location at the edge of the feeder, the important reductions of voltage deviations obtained with 3kW of batteries and the performance comparison among architectures.

\begin{table}[!htbp]

\scriptsize

\caption {Performance Indices for 3 kW storage deployments} \label{overloadBattery}


\begin{center}

\begin{tabular}{p{3.5cm}p{0.9cm}p{0.8cm}p{0.8cm}p{0.85cm}} 

\hline 

Storage deployment & Max VUF (\%) &Neutral Losses (kWh)& Phase Losses (kWh)& Max V drop/rise (\%) \\

\hline 

\hline 

No storage - Balanced & 1.7e-7 & 0.00 & 16.19 & 3.65\\

\hline 

No storage - Unbalanced & \textcolor{red}{1.30} & \textcolor{red}{3.29} &	\textcolor{red}{21.53} &	\textcolor{red}{8.30}\\

\hline

\hline

Architecture 1 - N0 & \textcolor{red}{1.26} & \textcolor{red}{3.28} & \textcolor{red}{21.52} & \textcolor{red}{8.04}\\

\hline

Architecture 1 - N5 & \textcolor{green}{0.87} & \textcolor{green}{1.78} & \textcolor{green}{18.65} & \textcolor{green}{5.98}\\

\hline

\hline

Architecture 2 - N0 & \textcolor{red}{1.28} & \textcolor{red}{3.28} & \textcolor{red}{21.52} & \textcolor{red}{8.07}\\

\hline

Architecture 2 - N5 & 1.01 & 2.17 & 19.90 & 6.97\\

\hline

\hline

Architecture 2 - N5 - No load shift & 1.08 & 2.41 & 20.09 & 7.18\\

\hline

\end{tabular}

\end{center}

\end{table}


The results in Tab.~\ref{overloadBattery} shows that locating the battery at the substation level has a very weak impact on the metrics compared to locating it at the edge of the grid. 
In terms of architecture comparison, we can see that architecture 1 outperforms architecture 2 in all scenarios considered. 
This is because the unbalance is concentrated in only one phase.
However, a unbalance not concentrated in one phase may be better balanced by architecture 3.

The comparison among the combined impact of phase balancing and load shifting and the case in which load is not shifted in time shows that the sensibility of the performance metrics to phase switching is dominant with respect to load shifting. In particular if we check the last two lines of Tab.~\ref{overloadBattery} we observe that the results do not vary much when only phase switching is allowed for the same aggregated capacity of battery storage, but both approaches provide considerable reductions with respect to the case without batteries. For instance, max VUF gets reduced from 1.3\% to 1.01\% and 1.08\% for the cases with and without load shifting respectively. This raises the interest of carrying out a more throughout sensibility analysis, which will be the subject of future work.

\section{Conclusion and Perspectives} 
\label{phaseSection5}
Distribution grids will increasingly face efficiency and quality of supply issues due to network unbalance, which is exacerbated by DGs and EVs. Nevertheless, this is an underestimated problem that already affects distribution grids as we demonstrate through the analysis of two case studies. The first case study is derived from data collected from Madeira Portugal, where the system operator statically allocates phases with the aim to minimize the impact of DG, but the network imbalance remains important. The second case study deals with behind the meter network unbalance in an EV charging testbed located at Caltech in Pasadena, California.

We characterize the negative impacts of network unbalance based on simulations of a radial distribution network using an OpenDSS software platform and the following efficiency/quality metrics: (a) VUF, (b) neutral losses and (c) line losses and (d) voltage drop/rise. 
Depending on network types and loading levels the imbalance indices most affected differ. 
We observe that DGs/EVs connected at farther end of a network cause significantly more imbalance compared to same amount connected closer to the feeder. This implies the correction of imbalance indices can be higher if storage is connected at the farther end of a radial distribution network.
%
We discuss three possible architectures for their integration to the distribution grid.
We rely on stylized simulations to 
show that significantly smaller sized storage compared to imbalance magnitude can considerably improve network imbalance metrics.
Furthermore, we show the sensitivity of the metrics to phase balancing is dominant with respect to load shifting. 

All these observations represent a valuable input for the development of storage control algorithms aimed to reduce network unbalance, in which we are currently working on.

\bibliographystyle{IEEEtran}
\bibliography{thesisch0}

\end{document}